\documentclass[letterpaper,twocolumn,superscriptaddress,showkeys]{revtex4}
\usepackage[latin1]{inputenc}
\usepackage{amsmath,dcolumn,graphicx,hyperref}
\hypersetup
{
    colorlinks = true, linkcolor = blue, citecolor = blue, urlcolor = blue,
    pdfauthor = {Desjardins-Proulx, Philippe},
}
\begin{document}

\title{A complex speciation-richness relationship in a simple neutral model}

\author{Philippe Desjardins-Proulx}
\email[E-mail: ]{philippe.d.proulx@gmail.com}
\affiliation{College of Engineering, University of Illinois at Chicago, USA.}

\author{Dominique Gravel}
\affiliation{Canada Research Chair on Terrestrial Ecosystems, Universit\'e du Qu\'ebec, Canada.}

\begin{abstract}
    Speciation is the ``elephant in the room'' of community ecology. As the ultimate source of biodiversity, its integration in ecology's theoretical corpus is necessary to understand community assembly. Yet, speciation is often completely ignored or stripped of its spatial dimension. Recent approaches based on network theory have allowed ecologists to effectively model complex landscapes. In this study, we use this framework to model allopatric and parapatric speciation in networks of communities. We focus on the relationship between speciation, richness, and the spatial structure of communities. We find a strong opposition between speciation and local richness, with speciation being more common in isolated communities and local richness being higher in more connected communities. Unlike previous models, we also find a transition to a positive relationship between speciation and local richness when dispersal is low and the number of communities is small. We use several measures of centrality to characterize the effect of network structure on diversity. The degree, the simplest measure of centrality, is the best predictor of local richness and speciation, although it loses some of its predictive power as connectivity grows. Our framework shows how a simple neutral model can be combined with network theory to reveal complex relationships between speciation, richness, and the spatial organization of populations.
\end{abstract}


\maketitle

\section{Introduction}

For a long time speciation was not part of community ecology's theoretical framework. MacArthur and Wilson's seminal work on island biogeography does mention speciation but their model and most of its inheritors ignored it completely \cite{mac67}. This is surprising given speciation's central role: ultimately, all species appear through speciation events. The importance of speciation to understand patterns of diversity was noted by Wallace in the 1850s \cite{wal55} and played a key role in the modern synthesis of evolutionary biology \cite{may42}. Fortunately, ecologists are increasingly aware of the importance of speciation, in part because of a growing interest in the influence of regional processes on local diversity \cite{ric08}. Recently Vellend argued that, while a great number of processes shape communities, they can be grouped in four classes: drift, dispersal, selection, and speciation \cite{vel09,vel10}. Recent theoretical models, such as those based on Hubbell's neutral theory \cite{hub01} or the Webworld \cite{cal98}, have also made speciation an important part of community ecology \cite{dro01,hub05b,eti07,odw10,kop10,ros10,mel10,ros11,eti11}. In particular, the neutral theory covers three of the four classes of processes described by Vellend, leaving only selection untouched \cite{hub01,vel10} in favor of a more tractable (some would say pragmatic) description of community dynamics \cite{wen12}. However, whereas drift and dispersal are well integrated in the neutral theory, the treatment of speciation remains dubious \cite{eti07,des12a}.

\begin{figure}[ht!]
    \centering\includegraphics[width=0.45\textwidth]{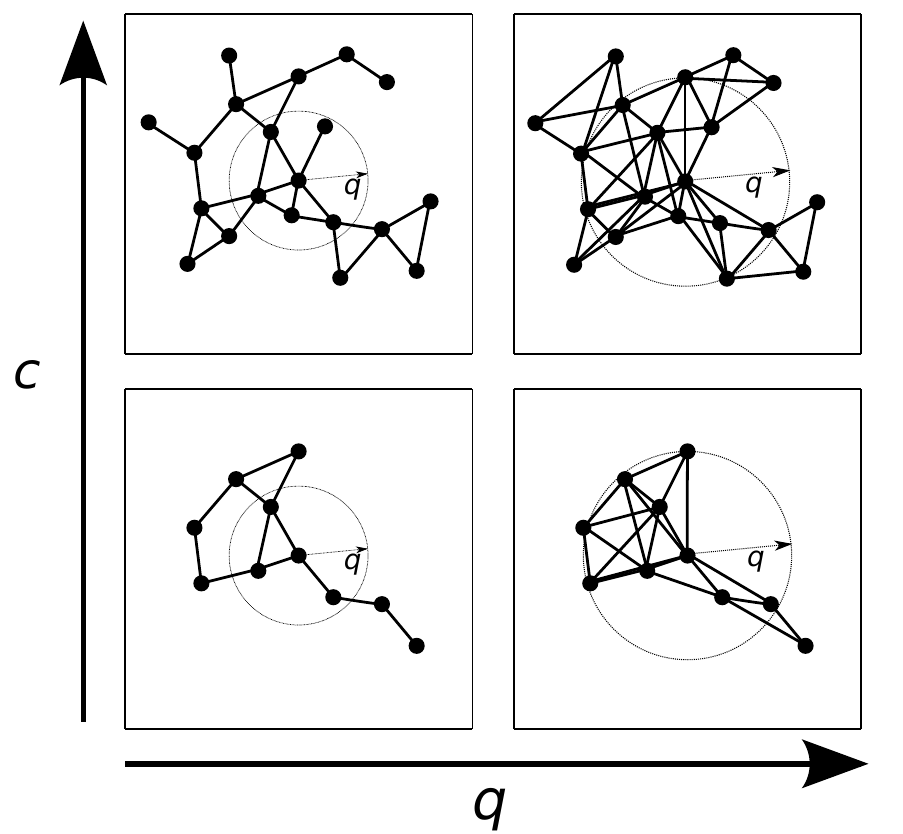}
    \caption
    {
        Four metacommunities represented by random geometric networks in the unit square $(x, y \in [0, 1])$. We define the metacommunity as the entire network of communities. Here the local communities (the vertices) are represented by black circles and the thick black lines (the edges) denote links by dispersal. Each of the $c$ vertices has a position in two-dimensional space and is linked to all vertices within some threshold Euclidean distance $q$, which can be seen as the dispersal range of the species. A community is a set of populations of different species. We define a population as the entire set of individuals of a given species in a given vertex. As $q$ increases, the number of links grows larger and local communities are less isolated. Similarly, the number of links per community also increases with $c$. While these networks are random, they exhibit locality, an important feature of real landscapes. Networks are well suited to distinguish populations within a species and thus to model speciation as a population process. Dispersal rates between connected communities are always low so we can assume the individuals of a given species in a given local community is a population in the strict sense \cite{ber02}. Within each local community, the populations of one or more species fluctuate by drift and dispersal in the exact same way as Hubbell's neutral model \cite{hub01}. Unlike the previous models by Hubbell \cite{hub01} and Economo and Keitt \cite{eco08}, we model speciation as a population process \cite{des12a}. Populations diverge through mutations and within-species selection. A population will undergoe speciation if it accumulates enough genetic differences. In this model, each local community offers a possibility of speciation, which is inhibited by the homogenizing effect of gene flow. The rate of speciation is determined by the number of populations in the metacommunity and how much inhibiting gene flow is present. Speciation is thus an emergent property of the metacommunity.
    }
    \label{fig:rgg}
\end{figure}

In community ecology, speciation is often reduced to a mutation leading instantaneously to a new species with a single individual. We refer to this modeling approach as ``speciation-as-a-mutation''. This is the approach of both Hubbell's original neural theory \cite{hub01} and Webworld \cite{cal98}. A notable exception is Rosindell's protracted speciation model, where speciation is modelled as a gradual process similar to the model we propose \cite{ros10,ros11,des12a}. Because the processes determining the fate of mutations (gene flow, selection, drift) have no effect on the mutation rate, population geneticists can simply obtain an estimate from field data and plug the mutation rate in equations \cite{cro70,vel03}. Speciation is different. The speciation rate is an emergent property of selection, drift, and dispersal processes \cite{coy04}. To put it another way: a mutation is a molecular phenomenon mostly unaffected by allele dynamics so it can be treated independently. Speciation on the other hand is a population process influenced by the structure and dynamics of populations, so it cannot be treated as a fixed rate. In particular, as gene flow tends to inhibit divergence, the spatial organization of a species' populations and their level of isolation will determine the likelihood of speciation \cite{coy04,ric08}. In this study we use networks of communities to move the neutral theory from ``speciation-as-a-mutation'' to ``speciation-as-a-population-process'' (Fig. \ref{fig:rgg}).

Spatial patterns of diversity are notoriously hard to study theoretically. Part of the problem lies in the lack of effective analytical methods for nontrivial spatial models \cite{epp10} (but see \cite{odw10}). Network theory provides tools to study patterns of connections and allows us to model almost any kind of spatial structure \cite{dal10}. Furthermore, theorists have developed algorithms to analyze various aspects of networks, making it an effective tool to extract information from highly complex structures \cite{new10}. For example, we use several algorithms to quantify how different types of centrality influence speciation and diversity. A network is simply defined by two sets: a set of vertices and a set of edges. In our case, the vertices represent local communities and the edges denote dispersal (Fig. \ref{fig:rgg}). The entire network forms the metacommunity. A spatial network combines the combinatorial properties of a network with a topological space in any number of dimensions \cite{kob94}. They have been used, among other things, to study networks on maps \cite{sed02,pen03,dal10} and the three-dimensional structure of molecules \cite{shi03}.

Economo and Keitt pioneered the use of spatial networks in metacommunity theory \cite{eco08,eco10}. They extended Hubbell's neutral theory to networks and studied how different topologies influence diversity. However, they kept the ``speciation-as-a-mutation'' model and did not use the network to account for the influence of isolation and gene flow on speciation \cite{eco08,eco10}. To introduce speciation in a realistic matter, we have to go beyond the ``speciation-as-a-mutation'' framework and treat it as a population-process inhibited by gene flow (Fig. \ref{fig:rgg}). Speciation modes are most often distinguished by their biogeography \cite{coy04}. Allopatric speciation occurs when the new species originates from a geographically isolated population, sympatric speciation is often defined as speciation without geographical isolation, and finally, parapatric speciation covers the middle ground between these two extremes \cite{coy04}. The relative importance of gene flow to speciation is still hostly debated \cite{nos08,joh10}. Nonetheless, speciation with little or no gene flow is still thought to be more common \cite{coy04,fit08,bol07}. To study the effect of speciation on diversity, we extend the framework of Economo and Keitt to a population-based speciation model \cite{des12a}. We show that treating speciation as a population-process inhibited by gene flow has profound consequences on the predicted patterns of diversity. We discover a complex relationship between speciation and local richness. When the number of local communities is small, we find a strong positive relationship: communities with more speciation events are also the ones with the highest richness. However, this relationship does not hold as the number of communities increases. Finally, we compare the effectiveness of different centrality measures as predictors of local richness and speciation.

\section{Methods}

\subsection{Metacommunity dynamics}

The metacommunity dynamics is similar to Hubbell's neutral model of biodiversity \cite{hub01}. We model the metacommunity as a network of $c$ local communities (Fig. \ref{fig:rgg}). The dynamics can be described in three steps \cite[Fig. 1]{des12a}. 1: For each time step an individual is selected and killed in each community, with all individuals having the same probability of being selected. 2: The individuals selected in step 1 are replaced either by dispersal or by local replacement. The probability of dispersal from vertex $x$ to vertex $y$ is given by the dispersal matrix $\mathbf{m}$. The dispersal matrix is built by giving all self-loops ($m_{xy}$ with $x = y$) a weight of 1, all connected communities a weight of $\omega$, and then dividing all values by the sum of the weights so each row will sum to 1. The rate of dispersal for a given vertex is roughly $l\omega$, with $l$ being the number of connected communities \cite{des12a}. In the case of dispersal from $x$ to $y$ ($x \not = y$), the new individual belongs to species $i$ with probability $N_{ix}/J_x$, with $N_{ix}$ being the population size of species $i$ in community $x$ and $J_x$ the total number of individuals in local community. Each individual belongs to a species and carry a genotype, either $a_xb_x$, $A_xb_x$, or $A_xB_x$, with $x$ being the community. We assume that migrants carry no mutations at the focal loci for the population into which they move so the haplotype is ignored and the new individual will always carry $a_yb_y$. This assumption implies that each population has its own unique path to speciation, an assumption made necessary by neutrality and the lack of environmental variables. For local replacement events, the new individual will belong to species $i$ with probability $N_{ix}/J_x$. However, the fitness of the haplotypes is used to determine the new individual's haplotype. One of the basic tenets of the neutral theory is ecological equivalence, so to introduce selection within the framework of neutral ecology the probability to pick an individual from one species has to ignore the internal genetic composition. Thus, our model is neutral when it comes to interspecific community dynamics but includes intraspecific selection in favor of new mutations leading to speciation. After the species is selected, we select the haplotype using the fitness $1.0$, $1 + s$, $(1 + s)^2$ for the three haplotypes respectively. When the haplotype is selected, $a_xb_x$ mutates to $A_xb_x$ and $A_xb_x$ to $A_xB_x$ with probability $\mu$. 3: Lastly, all populations with $A_xB_x$ fixed undergo speciation. In short, if all the individuals of a given species in a given vertex carry the $A_xB_x$ haplotype, they speciate. The individuals of the new species will carry $a_xb_x$ and a new path toward speciation is open. A more detailed description of the model can be found in Desjardins-Proulx and Gravel \cite{des12a}.

We study metacommunities with a varying number of vertices (local communities) $c$ and threshold values $q$. We generate the random geometric networks by randomly placing the vertices in the unit square and connecting all vertices within some Euclidean distance $q$ (Fig. \ref{fig:rgg}) \cite{sed02,pen03}. We generate random geometric networks until a connected one is found. This method might introduce a bias as many networks will be rejected when $q$ is small but it is necessary because the presence of disconnected components makes the analysis of networks notoriously difficult \cite{new10}. The number of communities $c$ vary from 5 to 125 with a fixed metacommunity size of $100 000$ individuals (i.e.: we have 5 communities of 20 000 individuals, or 10 of 10 000, or 25 of 4 000, and so on). In a previous study we found that global diversity was optimal around $s = 0.15$ and $\omega = 5e-4$ and we use these values unless otherwise noted \cite{des12a}. $\omega$ is a parameter used to create the dispersal matrix and can roughly be defined as the dispersal rate between two vertices \cite{des12a}. We also tried different values of $s$ to test the solidly of our results ($s = \{0.05, 0.10, 0.20, 0.30, 0.40\}$). We set the mutation rate $\mu$ to $1e-4$, a high but realistic value for eukaryotes \cite{dra98,gav04}. See Desjardins-Proulx and Gravel for details on the influence of selection, $\omega$, and the mutation rate on diversity \cite{des12a}. All simulations started with 20 species evenly distributed in the metacommunity and ran for 100 000 generations. The simulations were written in ANSI C99 and the code is available on github (\href{https://github.com/PhDP/origin}{https://github.com/PhDP/origin}).

\subsection{Centrality measures}

We explore the effect of five measures of centrality and importance on diversity \cite{new10}. The first is the degree of the vertex, which is the number of edges starting from the vertex plus the number of edges going into the vertex divided by two. The second measure is eigen-centrality. It assigns scores to vertices so that connections to high-scoring nodes are more important than connections to low-scoring vertices. Closeness centrality is the average geodesic distance between the vertex and all other vertices. Unlike the degree, which is only affected by the neighbors, closeness centrality depends on the global structure of the network. Betweenness centrality is the number of shortest paths from all vertices to all others that pass through the vertex. In short, if we compute the shortest paths for all pairs of vertices, how many times a vertex is present in these paths determine its betweenness centrality. Lastly, communicability centrality is the sum of closed walks of all lengths starting and ending at the vertex. We also studied the effect of clustering on diversity. Clustering is defined as the density of triangles in the network, or the probability that two neighbors of a vertex are themselves neighbors \cite{new10}. Clustering is expected to be fairly high in random geometric networks because all vertices have a position in space and this position is used to establish connections. For a vertex $v$, clustering is defined as: 

$$ C(v) = \frac{2T}{\deg(v)(\deg(v) - 1)}, $$ 

where $T$ is the number of triangles passing through vertex $v$ and $\deg(v)$ is the degree of $v$. We used the open-source Python library NetworkX to compute these centrality measures and the clustering coefficients \cite{hag08}.

\section{Results}

\subsection{Local patterns of diversity}

\begin{figure*}[ht!]
    \centering\includegraphics[width=1.0\textwidth]{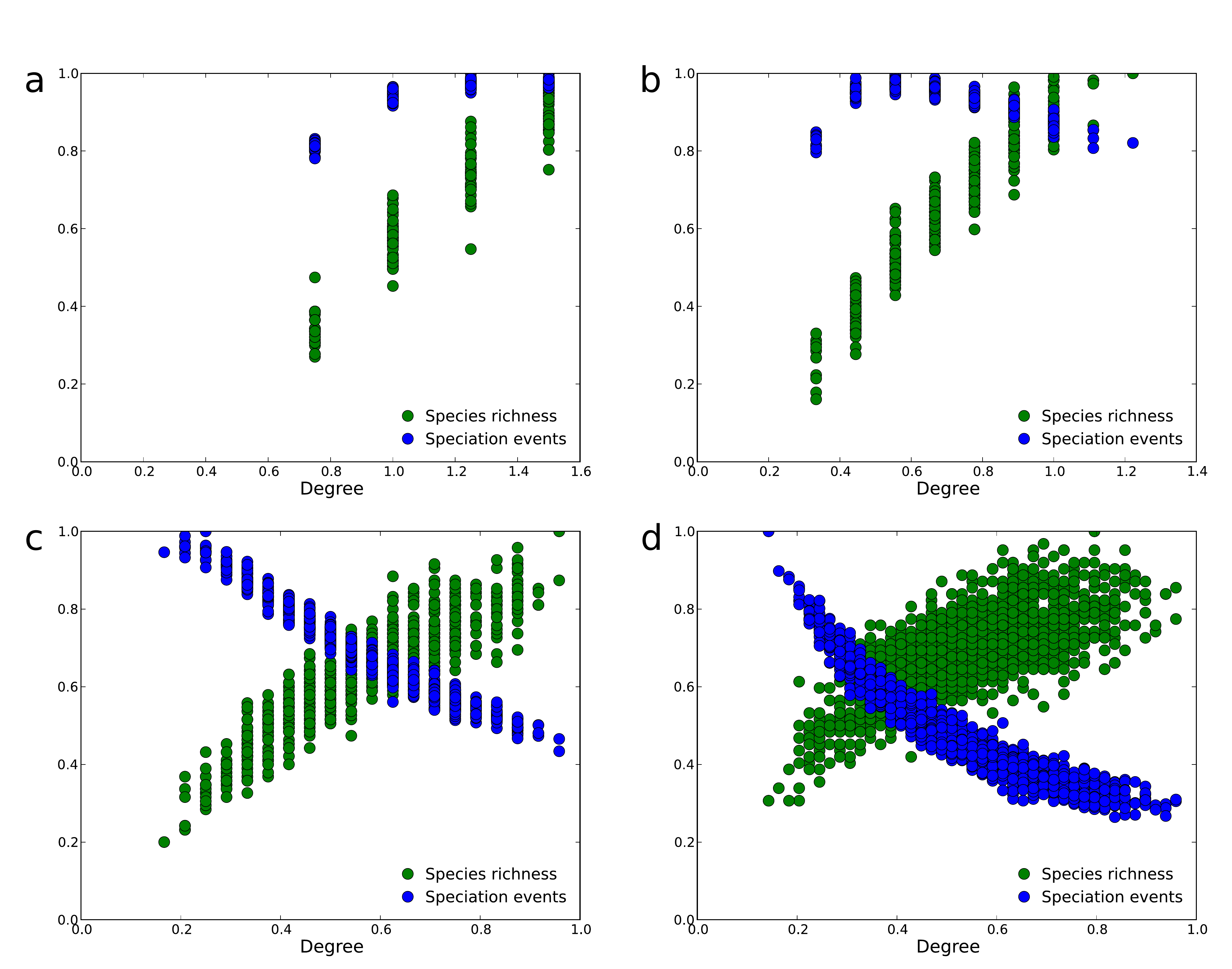}
    \caption
    {
        The relationship between degree centrality and local patterns of richness/speciation in metacommunities with 5 vertices (a), 10 vertices (b), 25 vertices (c) and 50 vertices (d). Species richness, the number of speciation events, and the vertex degree are all normalized. The points represent the vertices (local communities) at the end of the simulation and show how much local richness and how many speciation events were found according to their degree centrality. The average degree gets higher as the number of vertices in the metacommunity, denoted $c$, increases. In practice it means higher dispersal as $c$ increases. When the number of vertices is small, speciation and richness are positively correlated. With $c = 10$ we observe the transition between a positive relationship to a negative one (b). It is the only plot where communities with an intermediate degree have the most speciation events. For larger metacommunities with higher dispersal (c \& d), there is a clear negative relationship between speciation and local diversity. The models of Hubbell and Economo \cite{hub01,eco08} use a constant speciation rate per individual, so a similar plot would yield a flat line for speciation events (i.e.: it is unrelated to richness or dispersal). Selection was fixed at $s = 0.15$ and we used a threshold value of $q = 0.50$ to generate the random networks. We ran 32 simulations for each figure, leading to $32 \times c$ data points per figure.
    }
    \label{fig:spedivdeg}
\end{figure*}

\begin{figure*}[ht!]
    \centering\includegraphics[width=1.0\textwidth]{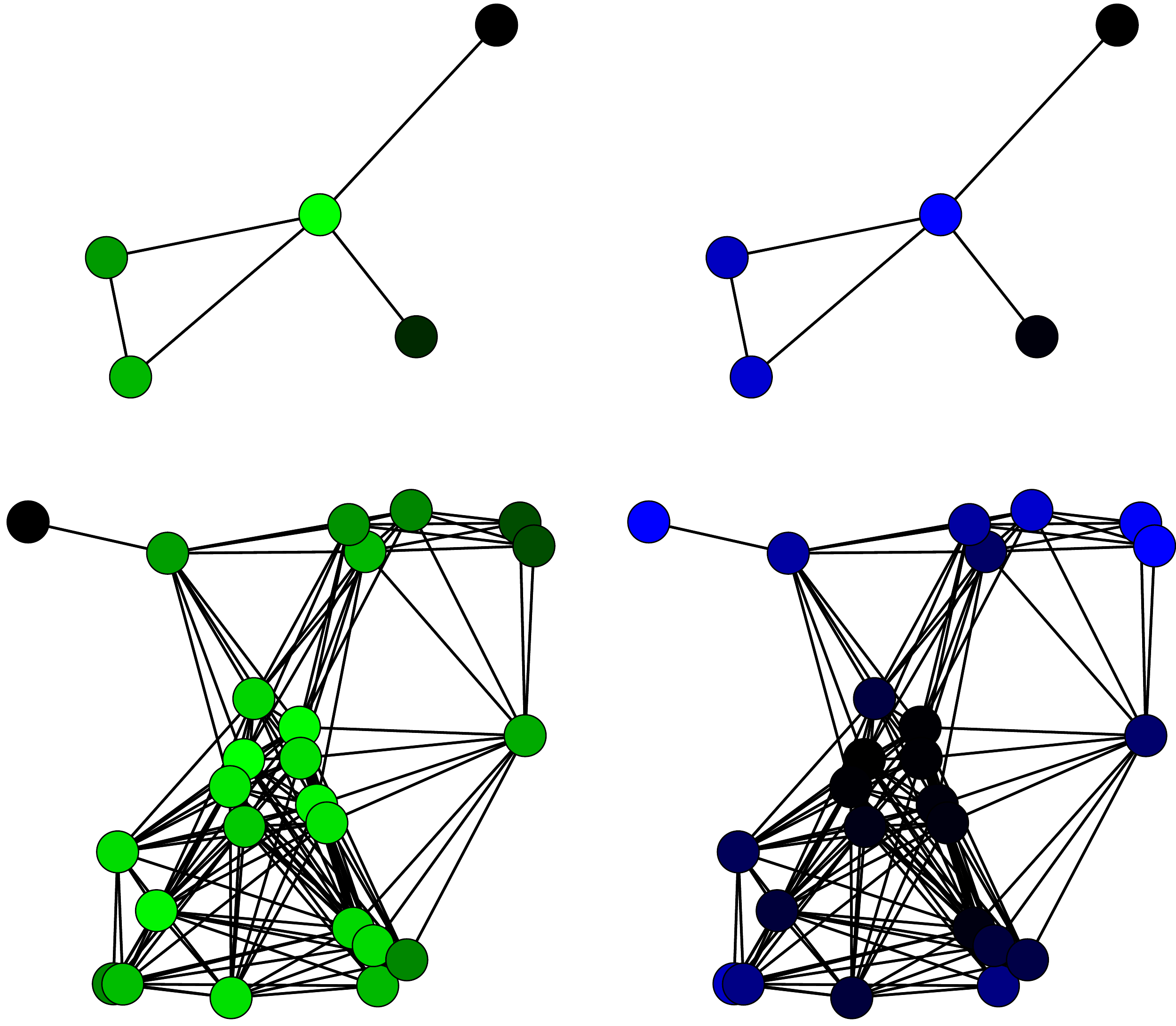}
    \caption
    {
        Local richness and speciation in networks of communities for two simulations. On top: a simulation with 5 communities and at the bottom: a simulation with 25 communities. On the left, communities change from black to green as local richness increases and on the right, communities change from black to blue as the number of speciation events increases. For the metacommunity with 5 communities, richness and speciation events are strongly correlated. The most connected communities support more species and more speciation events. With 25 communities the opposite is true. The communities with the most speciation events are far from the geodesic center, where local richness is higher.
    }
    \label{fig:ex}
\end{figure*}

\begin{figure}[ht!]
    \centering\includegraphics[width=0.5\textwidth]{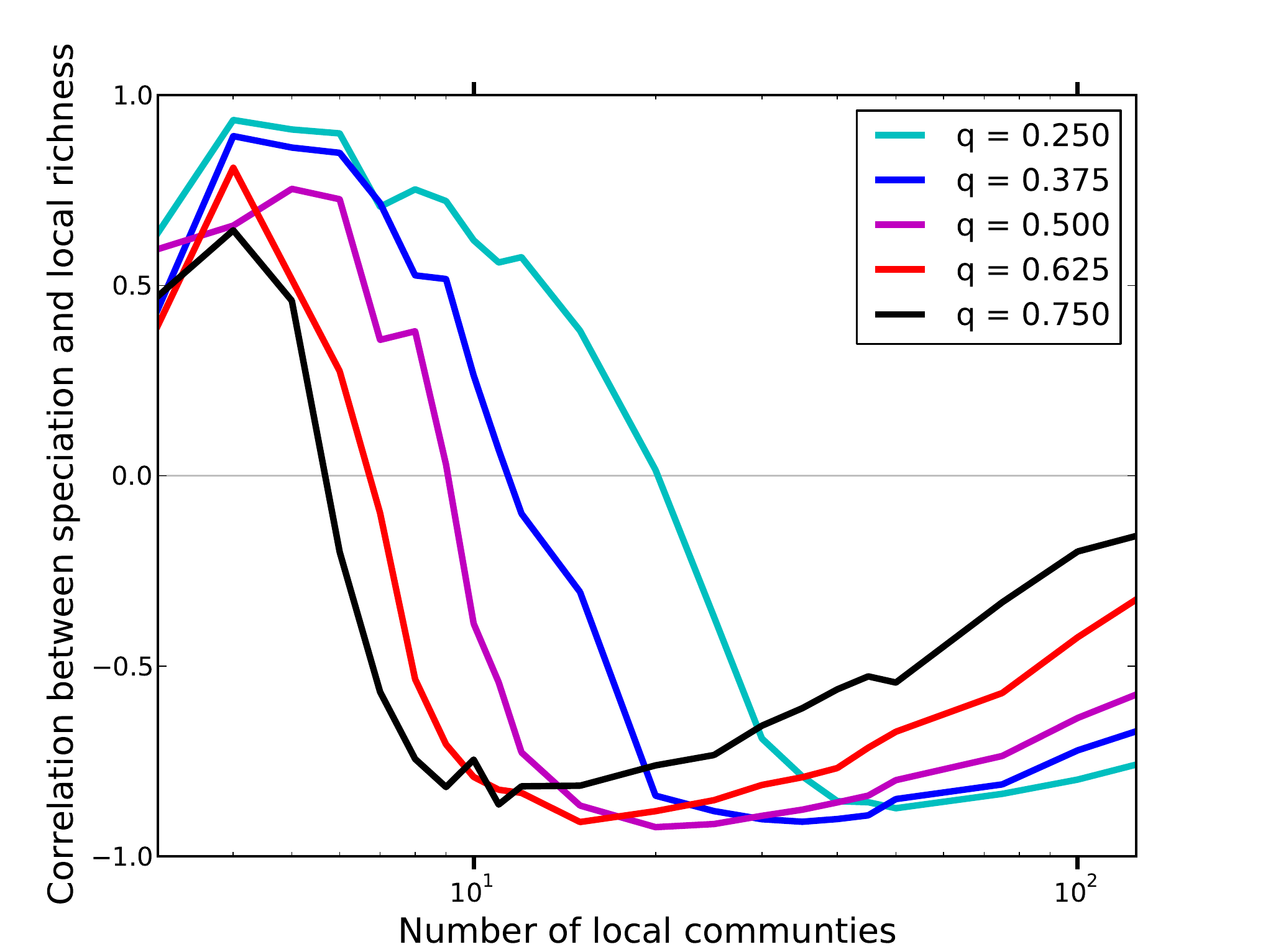}
    \caption
    {
        The relationship between speciation and diversity with an increasing number of local communities (total number of individuals in the metacommunity fixed at 100 000). Between $c = 3$ and $c = 8$, there is a positive relationship between them. As $c$ approaches 10, the correlation drops and quickly becomes negative. It reaches a peak around $c = 25$, where the correlation starts to grow weaker. This complex relationship is a direct consequence of the ``speciation-as-a-population-process'' framework. Because the number of links per community is low with $c < 10$, communities with more species have more opportunities for speciation without being crushed by inhibiting gene flow. As $c$ increases, a larger number of individuals are migrants. They not only inhibit speciation but also take precious space. Populations fluctuate randomly and eventually become extinct. As the size of local communities grows smaller and the number of migrants grows, populations are more likely to become extinct before having the chance to speciate. These simulations used a within-species selection coefficient of $s = 0.15$ for the mutations leading to speciation. We tried different values $s = \{0.05, 0.10, 0.20, 0.30, 0.40\}$ and found similar patterns. With a purely neutral model ($s = 0.0$), the metacommunities could not support more than a few species \cite{des12a}.
    }
    \label{fig:trans}
\end{figure}

We first analyzed diversity on a vertex-by-vertex basis to understand the effect of network structure on local diversity. For all vertices we counted the number of speciation events, local richness at the end of the simulation, local and global extinctions events, and various measures of centrality. We compared metacommunities with 100 000 individuals divided into $c$ local communities and used a threshold value of $q$ to generate the random geometric network (see Fig. \ref{fig:rgg}). Species richness and the number of speciation events show strong positive correlations when $c$ is small and strong negative correlations when $c$ is large (Fig. \ref{fig:spedivdeg}). An interesting trade-off occurs: communities that are more connected (high $c$) can host more species, which means more opportunities for speciation. On the other hand, they host more species because of greater dispersal, and greater dispersal means greater inhibiting gene flow for speciation (Fig. \ref{fig:ex}). When $c$ is fairly large, the communities with a greater number of links support more species than isolated communities, but the effect of gene flow is so strong that speciation is very difficult. Furthermore, because so many individuals come from migration events, it becomes hard for local populations to diverge enough to speciate. This result is coherent with known patterns of diversity and speciation \cite{dia72,gav09}. However, when $c$ is small the total number of links, even in well-connected communities, is also small. The inhibiting effect of gene flow is still present but communities with more connections and more species will still witness more speciation events (Fig. \ref{fig:trans}).

Then we investigated the relationship between local diversity and several properties of the spatial network. In particular, we studied how different measures of centrality could be used to predict local richness and speciation events. The degree of a vertex (local community) is the crudest measure of centrality. While it gives information about how many communities are linked to the community of interest, it says nothing about the influence of the network's structure. Still, for all metacommunity sizes and all values of $q$, the degree shows the highest correlation with species richness and speciation (Table \ref{table:corr}). The correlation between the degree and species richness is close to one in many cases and performs poorly only for very high $c$ and $q$, where the number of species in the metacommunity becomes small. Closeness centrality is a measure of the average geodesic distance (the length of the shortest path in the network \cite{dij59}) between a vertex and all other vertices. Closeness centrality's ability to predict local diversity is thus related to the metacommunity size and global structure. Closeness centrality is a worse predictor than the degree and achieves better result for small $c$ (Table \ref{table:corr}). Closeness centrality is affected by all vertices, even those that are very far and unlikely to have any impact on the vertex. This is why closeness centrality performs better for small metacommunities: all communities are close so the measure cannot be biased by distant vertices. Eigen-centrality is an interesting alternative to the degree and closeness centrality. Unlike the first, it is affected by more than just neighbors but unlike the second far off vertices will have little effect on it. Overall, eigen-centrality is very similar to the degree but perform a little worst on larger communities, again showing the disproportionate effect of neighbors and the small effect of the overall structure of the network (Table \ref{table:corr}). Unsurprisingly, betweenness centrality performs rather poorly compared to the other measures (Table \ref{table:corr}). Local extinction is always strongly correlated with diversity ($r > 0.80$) and global extinction is strongly correlated with speciation ($r > 0.80$). Patterns of local and global extinctions closely follow the patterns of local richness and speciation, with a strong positive relationship with small $c$ and a transition to a negative relationship as $c$ increases.

\begin{table*}
    \centering\begin{tabular}{|l||ccccc||ccccc|}
    \hline
    & \multicolumn{5}{c||}{Speciation}                & \multicolumn{5}{c|}{Local diversity} \\
    \hline
    With $q = 0.25$ &   5  &  10  &  25  &  50  & 125  &   5  &  10  &  25  &  50  & 125  \\
    \hline
    Degree          & 0.87 & 0.64 &-0.49 &-0.97 &-0.94 & 0.97 & 0.95 & 0.92 & 0.88 & 0.78 \\
    Eigen           & 0.87 & 0.53 &-0.52 &-0.76 &-0.73 & 0.97 & 0.84 & 0.69 & 0.68 & 0.63 \\
    Closeness       & 0.81 & 0.53 &-0.28 &-0.73 &-0.82 & 0.95 & 0.85 & 0.57 & 0.73 & 0.71 \\
    Comm.           & 0.81 & 0.59 &-0.56 &-0.80 &-0.65 & 0.92 & 0.90 & 0.82 & 0.69 & 0.57 \\
    Between.        & 0.68 & 0.38 &-0.22 &-0.40 &-0.57 & 0.82 & 0.70 & 0.38 & 0.41 & 0.53 \\
    \hline
    \hline
    With $q = 0.50$ &   5  &  10  &  25  &  50  & 125  &   5  &  10  &  25  &  50  & 125  \\
    \hline
    Degree          & 0.85 &-0.56 &-0.98 &-0.95 &-0.95 & 0.96 & 0.96 & 0.90 & 0.82 & 0.60 \\
    Eigen           & 0.88 &-0.50 &-0.93 &-0.93 &-0.94 & 0.95 & 0.92 & 0.86 & 0.79 & 0.59 \\
    Closeness       & 0.79 &-0.56 &-0.95 &-0.93 &-0.94 & 0.93 & 0.93 & 0.88 & 0.81 & 0.59 \\
    Comm.           & 0.86 &-0.56 &-0.91 &-0.88 &-0.90 & 0.95 & 0.92 & 0.83 & 0.76 & 0.56 \\
    Between.        & 0.54 &-0.57 &-0.67 &-0.74 &-0.79 & 0.74 & 0.68 & 0.62 & 0.66 & 0.49 \\
    \hline
    \hline
    With $q = 0.75$ &   5  &  10  &  25  &  50  & 125  &   5  &  10  &  25  &  50  & 125  \\
    \hline
    Degree          & 0.51 &-0.93 &-0.95 &-0.92 &-0.85 & 0.94 & 0.90 & 0.77 & 0.59 & 0.11 \\
    Eigen           & 0.56 &-0.90 &-0.95 &-0.92 &-0.86 & 0.93 & 0.89 & 0.76 & 0.55 & 0.11 \\
    Closeness       & 0.47 &-0.94 &-0.93 &-0.87 &-0.82 & 0.94 & 0.89 & 0.75 & 0.54 & 0.11 \\
    Comm.           & 0.44 &-0.91 &-0.94 &-0.90 &-0.84 & 0.75 & 0.89 & 0.75 & 0.54 & 0.10 \\
    Between.        & 0.29 &-0.81 &-0.80 &-0.81 &-0.73 & 0.84 & 0.70 & 0.64 & 0.48 & 0.10 \\
    \hline
    \end{tabular}
    \caption{Correlations between centrality measures and patterns of speciation and local richness. Degree-centrality outperforms closeness centrality for all combinations of $q$ and $c$, degree-centrality outperforms eigen-centrality in 28 out of 30 cases, and eigen-centrality beats closeness centrality in about the same number of cases. Betweenness centrality has the worst performance. The results for $c = 125$ and $q = 0.75$ might seem puzzling but are actually simple to explain: at this point the metacommunity supports only a few species, often only one, because of small local community size and the crushing effect of gene flow. In short, the measures of centrality are bad at predicting diversity and speciation because there is very little of either.}
    \label{table:corr}
\end{table*}

\subsection{Global patterns of diversity}

\begin{figure}[ht!]
    \centering\includegraphics[width=0.5\textwidth]{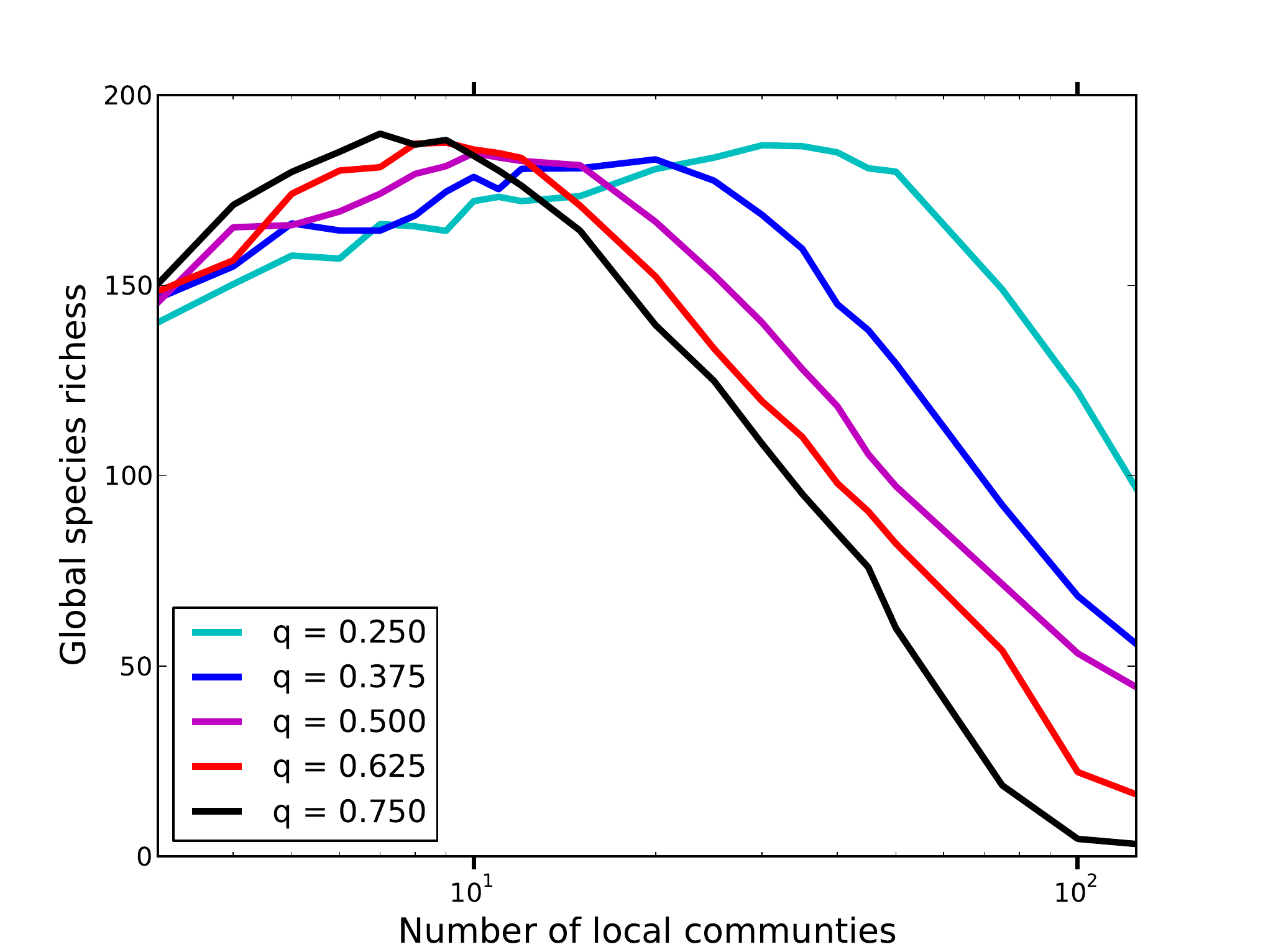}
    \caption
    {
        The relationship between global species richness and the number of vertices $c$ for various threshold distances $q$. Global diversity increases with higher connectivity and dispersal for 5 and 10 local communities. The effect of dispersal on global diversity is influenced by $c$. With 25 communities, richness at $q = 0.75$ is only 60\% of the diversity with $q = 0.25$. With 50 communities, it is 30\%, and with 125 communities: only 4\%. Communities get smaller with increasing $c$ and have more opportunities to form links. Thus, a much larger proportion of individuals are migrants, making speciation very hard to achieve. Several metacommunities with $c = 125$ and $q = 0.75$ had a single species at the end of simulations.
    }
    \label{fig:global}
\end{figure}

We investigated the effect of the network features such as the number of local communities $c$ and the radius $q$ used to generate the network. We find that global species richness is strongly affected by both $c$ and $q$ (Fig. \ref{fig:global}). For low values of $c$, diversity increases slightly with $q$. However, $q$ strongly reduces diversity for communities with high $c$. Because the dynamics is neutral and local communities with $n = 125$ can only support 800 individuals, a population is much less likely to stay long enough in a local community to undergo speciation. For $c = 5$ the species richness is roughly unaffected by $q$, increasing from 158 species with $q = 0.25$ to 182 with $q = 0.75$. This is not surprising since the average number of links is about equal (3.2 for $q = 0.25$ and 4.2 for $q = 0.75$). The effect become more pronounced as $c$ grows larger. For $c = 25$ the average number of links increases markedly from 5 ($q = 0.25$) to 20 ($q = 0.75$), and species richness decreases from 186 to 125. The effect of $q$ becomes evident at $c = 125$, where the average number of links jumps from 20 to 100, and diversity crashes from 100 to only 4 species. In other words, in a neutral model with allopatric/parapatric speciation, the dispersal range will have a strong inhibiting effect on speciation especially if the metacommunities are divided into a great number of communities that can only support a few individuals. These results highlight the complex role played by connectivity on diversity when the effect of gene flow on speciation is considered. On one hand connectivity inhibits speciation by increasing gene flow. On the other hand it promotes diversity by increasing dispersal. We show that an increase in connectivity will increase diversity if the metacommunity is divided into a few large communities and inhibit diversity as the number of communities increases.

We then took a closer look at the relationship between dispersal (threshold radius $q$) and diversity when the number of communities $c$ is held constant. $q$ increases the number of links and thus dispersal but its impact on the number of links depends on the number of communities and the formation of clusters. We studied the effect of dispersal on diversity by comparing metacommunities with the same number of communities $c$ and generated with the same threshold value $q$. Because the networks are randomly generated, the average number of links per community will vary even if $c$ and $q$ are fixed. The correlation between the number of links and global diversity varies greatly. For example, with $n = 5$ the correlation is very strong (r = 0.85), regardless of the value of $q$: more links equals higher diversity. On the other hand the correlation is almost nonexistent for $c = 125$. As the metacommunity grows larger, other structural characteristics of the networks play a greater role on diversity than the total number of links. Lastly, we explored the effect of clustering on species richness. Clustering is a measure of the tendency of vertices to form groups. High clustering provides more opportunities for dispersal but also decreases the number of isolated vertices. The correlation between the clustering coefficient and species richness decreases with both $c$ and $q$. The correlation is strongly positive for $c = 5$ with $r > 0.75$ for all values of $q$. The correlation decreases sharply with $q$ for $c = 10$: from 0.60 to 0.32 and -0.36 and with $c = 25$ from -0.11 to -0.30 and -0.58 for $q = 0.25, 0.50, 0.75$, respectively. It shows that, as $c$ and $q$ grows, clustering will tend to inhibit speciation enough to have a negative effect on diversity.

\section{Discussion}

Community ecology is about drift, dispersal, speciation and selection \cite{vel10}. Selection is a central component of community dynamics \cite{gra06} but if we want to understand its impact on spatial patterns of biodiversity we first need a solid reference template \cite{ros11b}. Neutral models reveal the spatial structure of biodiversity expected from the combined effects of dispersal, drift, and speciation in the absence of selection. Despite its simplicity, our model generates clear predictions on the relationship between richness and speciation. It explains patterns of interest to both community ecologists and evolutionary biologists by integrating speciation as a population process. Our model predicts that isolation reduces diversity, a well-known pattern in biogeography \cite{dia72}. It also predicts that isolation will stimulate speciation, a well-known pattern in speciation theory \cite{gav09}. Rosindell and Phillimore used a similar model and also found a negative relationship between speciation and local richness \cite{ros11}. The twist, in our model, is that this pattern is only true in some cases. When we model the metacommunity as a network with only a few communities, the most connected communities are both more diverse and witness more speciation events. In these cases, more diversity means more opportunities for speciation and the inhibiting effect of gene flow is not strong enough to counter the greater number of opportunities. Our models reveals a complex relationship between local richness, speciation, and isolation: one in which the spatial organization of communities and the strength of dispersal have the power to influence the speciation-richness relationship.

In our model the degree is a better predictor of diversity and speciation than the more sophisticated measures of centrality. The degree does not take into account the overall structure of the metacommunity, it is only determined by the number of neighbors. The fact that it is a better predictor than eigen-centrality and closeness centrality, which are both influenced by the overall structure of the community, shows that the behavior of our model is mostly driven by small scale patterns. It remains to be seen if this result is an artifact of ecological equivalence or a feature of real communities.

The neutral theory is often seen as weak when it comes to predictions related to speciation \cite{eti07,eti11}. In both Hubbell's model \cite{hub01} and its spatially-explicit counterpart \cite{eco08,eco10}, the rate of speciation is directly related to the number of individuals. Thus, it is not affected in one way or another by gene flow or the structure of the metacommunity. It is constant in a given community, regardless of the number of species present or the strength of dispersal. Our framework solves this problem by replacing the ``speciation-as-a-mutation'' approach with a speciation model based on populations \cite{des12a}. Our model predicts a negative relationship between diversity and speciation but it also predicts a positive relationship in some cases: namely when dispersal is weak. The exact relationship between diversity and speciation is complicated and not very well understood, especially from a mechanistic perspective. Emerson and Kolm did found a positive relationship between diversity and endemism, which could be interpreted as a rough index of speciation \cite{eme05}. They argue that diversity begets diversity: more species means more opportunities for other species to invade or speciate \cite{erw05}. Our model does predict a positive relationship in some cases but for different reasons. Cadena et al. and Witt and Maliakal-Witt \cite{cad05,wit07} suggest that species richness and endemism are positively correlated because of a mutual dependence on life spans \cite{cad05}. This is very similar to our model's prediction. When the metacommunity is divided into a few communities of large size, the gene flow is not strong enough to inhibit speciation and the greater local richness will offer more opportunities for speciation. The interesting twist in our framework is that this relationship will be reversed with greater gene flow and smaller communities. This prediction could be tested by comparing patterns of speciation and diversity in sets of islands of various sizes and connectivity.

One of the study's main limitation is arguably the neutral assumption. Our previous work suggests that neutral ecology is hard to reconcile with parapatric and allopatric speciation \cite{des12a}, requiring an uncomfortable compromise in the form of ecological equivalence at the species level and within-species selection. Some questions related to the relationship between speciation and richness will require the integration of adaptation and niches. For example, Emerson and Kolm's hypothesis that diversity stimulates speciation can only be tested in a trophic model \cite{gra11b}. Adaptive radiations and ecological speciation can easily define communities \cite{sch00,gil04c} and, because of their explosive nature, they might be more sensitive to the overall structure of the metacommunity. Also, some network structures are known to inhibit selection while others will stimulate it \cite{now06}. Another limitation is the simplicity of the speciation model used. Research on speciation genes are starting to yield fruits but we still do not have a clear picture of the most common road to speciation \cite{nos11b}, making it difficult to model the speciation process with precision. Our model is a simple attempt to include the most essential features of speciation: it is a population-process driven by the accumulation of genetic differences. Also, Hubbell's neutral theory is fundamentally similar to Moran processes and, for a given population, our model behaves as a Moran process with a fluctuating population \cite{mor62}. Coalescent theory and diffusion approximations are based on the same mathematical foundation, making it easy to relate our model to well-established evolutionary theory \cite{ewe04}. On the other hand, we make some unrealistic assumptions. Ecological speciation, in particular, cannot be modeled in a neutral framework. We had to assume that a road to speciation was always open, an assumption that greatly overestimates the rate of speciation. We also assumed a single mutation was enough to speciate. Further research should explore how relaxing the assumption of inter-specific ecological equivalence affects patterns of spatial diversity under allopatric and parapatric speciation.

Despite the limitations of neutrality, it provides a simple null hypothesis and could serve as a point of comparison with more realistic models of speciation in space. Theoretical community ecologists have mostly ignored the importance of space to model speciation. While recent developments have increased the visibility of speciation in community ecology, its treatment is often disconnected from speciation theory, making it difficult to unify ecology and evolution. Our approach to speciation is very simple but it captures many of the most important aspects of speciation: it is a population process often inhibited by gene flow and it cannot be treated as a simple mutation. Our work suggests that network theory, a promising framework to study patterns of diversity in space \cite{dal10}, can be used to integrate speciation in a more realistic matter and creates a bridge between community ecology and speciation theory.

\section{Acknowledgements}

We thank Hedvig Nenz\'en and two anonymous reviewers for her helpful comments on an earlier version of this manuscript. This work was supported by research grants from NSERC and the Canada Research Chair program to DG.

\bibliographystyle{plain}
\bibliography{/home/phdp/work/manuscripts/phdp}

\end{document}